\documentclass{iaus}
\usepackage{graphicx}

\title[S262.~~The interacting binary V393 Scorpii: another clue for 
Double Periodic Variables]
{The interacting binary V\,393 Scorpii: another clue for 
Double Periodic Variables}

\author[Mennickent et al.]
{Ronald Mennickent$^1$, Zbigniew Ko{\l}aczkowski$^{1,2}$,  Gojko Djurasevic$^3$, Gabriela Michalska$^1$
 \and Daniela Barr\'ia$^1$}

\affiliation{$^1$Departamento de Astronom\'{\i}a, Universidad de Concepci\'on, Chile \\ 
email: {\tt rmennick@udec.cl} \\[\affilskip]
$^2$Instytut Astronomiczny Uniwersytetu Wroclawskiego, Wroclaw, Poland \\[\affilskip]
$^3$Astronomical Observatory, Belgrade, Serbia 
 }

\pubyear{2009}
\volume{262}
\pagerange{...}
\jname{Stellar Populations-Planning for the Next Decade}
\editors{A.C. Editor, B.D. Editor \& C.E. Editor, eds.}
\begin{document}

\maketitle

\begin{abstract}
We give a brief report on spectroscopic properties of V\,393 Scorpii. H$\alpha$ emission and shape and radial velocity of  He\,I\,5875 are modulated with the long cycle. The long cycle is explained as a relaxation cycle in the circumprimary disc,  that cumulates the mass transferred  from the donor until certain instability produces disc depletion.

\end{abstract}

\firstsection 
       
\section{Introduction}

After the discovery of a long periodicity in the V\,393 Sco ASAS light curve by \cite[Pilecki \& Szczygiel (2007)] {2007IBVS.5768....1P}, we realized that this star is a bright Galactic member of the new class of interacting binaries Double Periodic Variables (DPVs), that are characterized by two closely related photometric periodicities (\cite[Mennickent et al. 2003] {2003A&A...399L..47M}, \cite[2009c]{2009arXiv0908.3900M}). They have been interpreted as intermediate mass semi-detached binaries experiencing cyclic mass loss into the interstellar medium (\cite[Mennickent et al. 2008]{2008MNRAS.389.1605M}, \cite[Mennickent \& Ko{\l}aczkowski 2009 a,b,c]{2009RMxAC..35..166M, 2009arXiv0904.1539M, 2009arXiv0908.3900M}). The mechanisms for cyclic mass loss still is a matter of research, and a study of V393 Sco could yield  important insigths on the DPV phenomenon. 

\begin{figure}[b]
\begin{center}
 \includegraphics[width=2.6in]{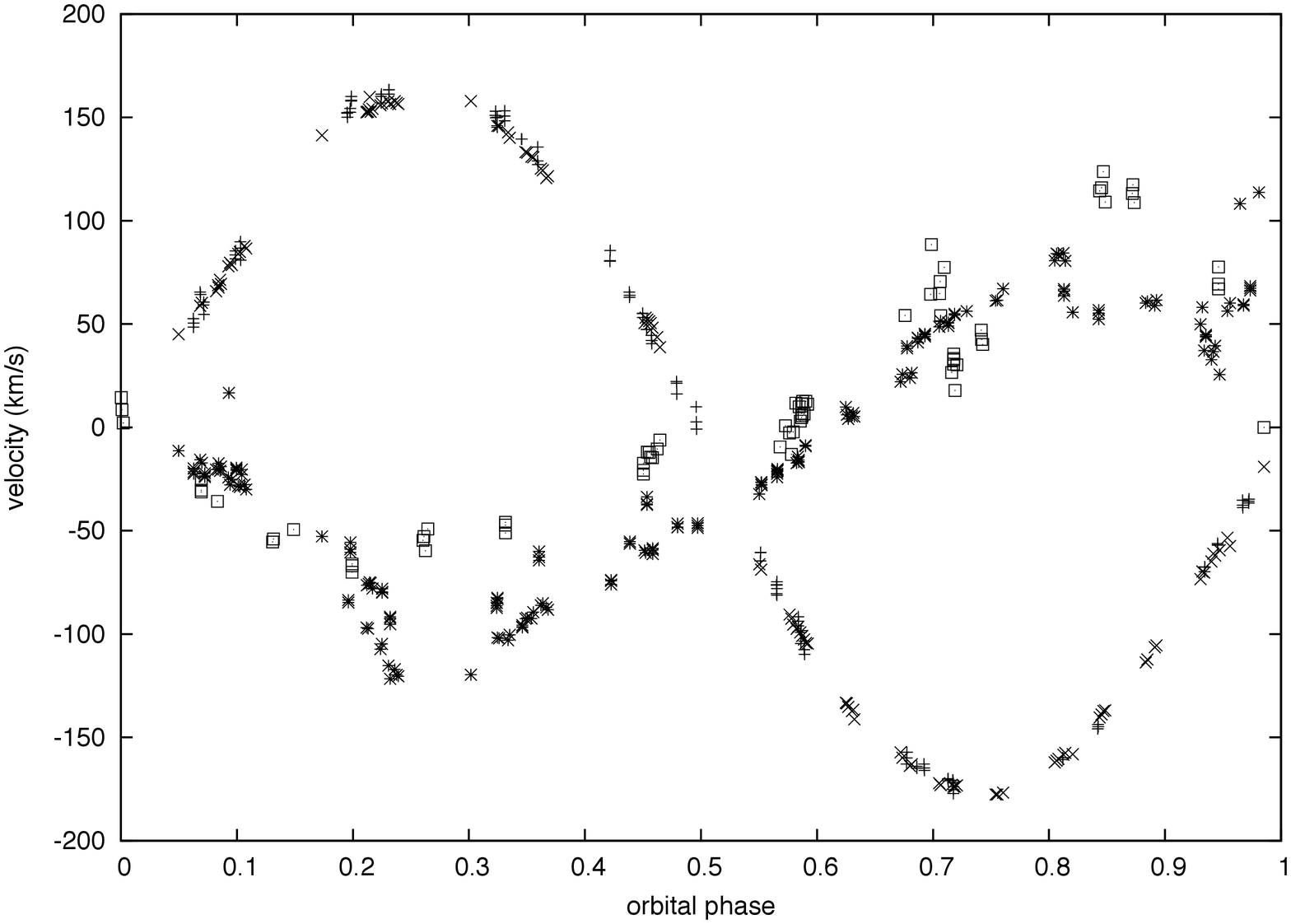} 
  \includegraphics[width=2.6in]{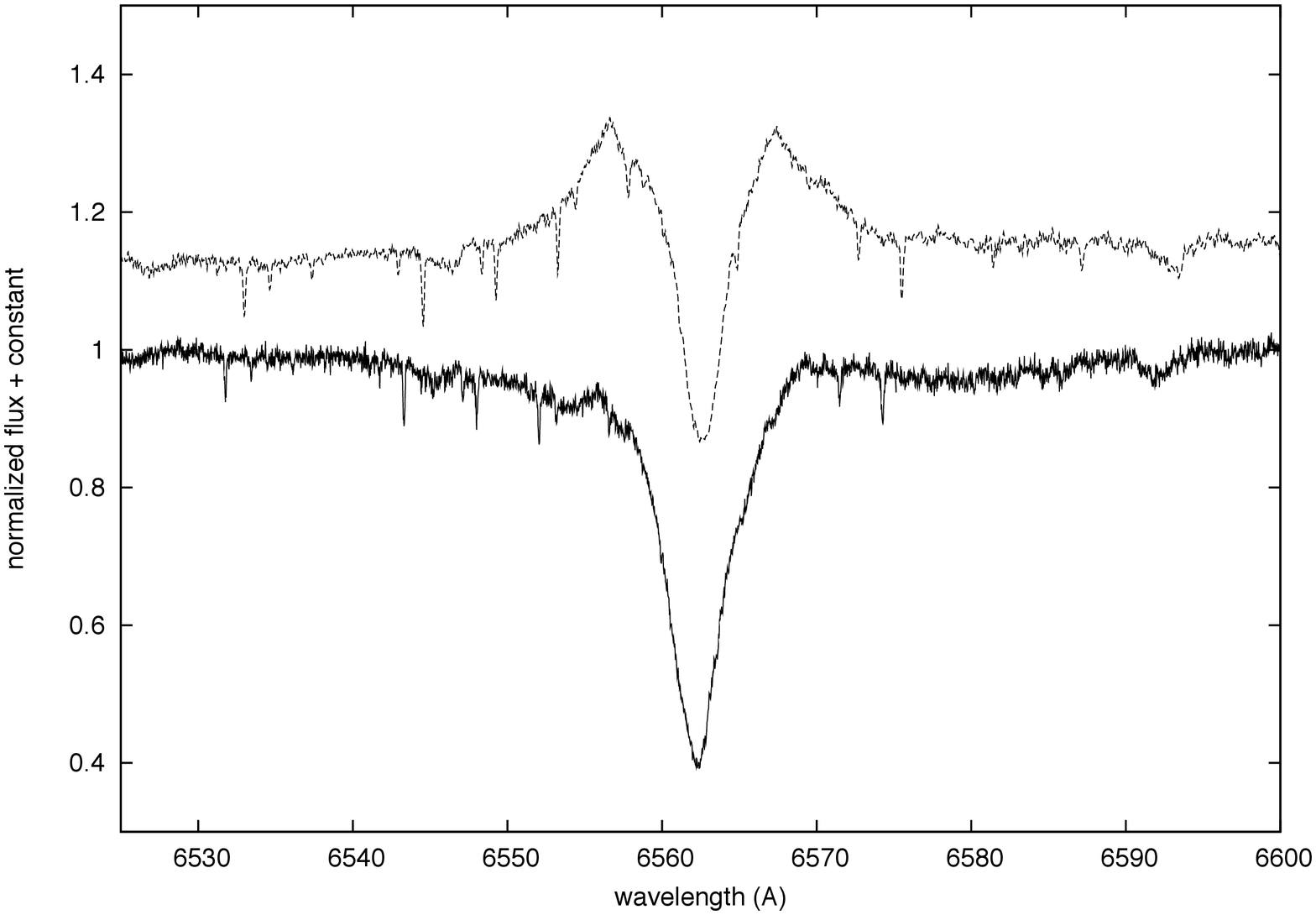}
 \caption{(Left) Radial velocities of the donor star yield $K_{2}$ = 169.7 $\pm$ 0.6 km/s (crosses and pluses) whereas He\,I\,5875 radial velocities at different long cycle phases, 0.2 $< \Phi_{l} \leq $ 0.8 (asterisks) and 0.8 $< \Phi_{l} \leq $  0.2 (squares), are peculiar.  Orbital ephemeris is from Kreiner (2004). (Right) H$\alpha$ emission at  $\Phi_{l}$= 0.94 ($\Phi_{o}$= 0.00, above) and  $\Phi_{l}$= 0.57 ($\Phi_{o}$= 0.97, down). }
   \label{fig1}
\end{center}
\end{figure}

\section{Observations}

During the last two years we have monitored V\,393 Sco with several high resolution spectrographs (e.g. CORALIE, UVES, FEROS, CRIRES) covering several orbital cycles and a couple of long cycles. We have also used the published ASAS V-band light curve and a special code (\cite[Djurasevic et al. 2008]{2008AJ....136..767D}, \cite[2009]{2009MNRAS.396.1553D}) for modeling the system. 
 
\section{Results}

Spectral lines of the donor are clearly visible and follow the orbital motion without interference during the long cycle, but the gainer (primary) is apparently (at least partially) hidden by a circumprimary disc  and He\,I absorption lines follow peculiar motions during orbital/long cycles (Fig.\,1). 
The H$\alpha$ double emission indicates an outer optically thin part of the circumprimary disc.  A Feros spectrum obtained at orbital phase 0.00 and long phase 0.94 shows H$\alpha$ emission much stronger than a UVES spectrum taken at orbital phase 0.97 and long phase 0.57 (Fig.\,1). This implies that the H$\alpha$ emission (eventually the disc) is larger on the long cycle maximum. The central absorption deeper than the continuum at this epoch  suggests that the disc is eclipsed by the donor.  
If we interpret the He\,I absorption lines as produced in a deeper, optically thicker disc region, their RVs indicate strong changes in the structure of this region during the long cycle. The presence of strong line absorption wings at specific phases at the He\,I 1083.3 nm line indicates mass exchange/loss in the system.   The strong modulation of the FWHM for the He\,I\,5875 line during long cycle could  reflect a rotational instability of the circumprimary disc (\cite[Mennickent et al. 2009c]{2009arXiv0908.3900M}). He\,I\,6678 is strongly influenced by emission shoulders and does not show this behavior. A comparison with a grid of spectra from  NLTE atmosphere models provided by Dr. Ewa Niemczura yields $T_{2}$= 7900 $\pm$ 100 $K$ and $\log\,g_{2}$= 3.0 $\pm$ 0.1. It is difficult to estimate the mass ratio, since the helium lines, that could be indicators of the primary motion, do not show a consistent value for $K_{1}$.  However, the best fit to the light curve of V\,393 Sco always requires the presence of a circumprimary disc under reasonable assumptions. Using $q$= 0.25 and $q$= 0.41 as trials we obtain $T_{1}$= 16910 $\pm$ 150 $K$ and $T_{1}$= 17400 $\pm$ 300 $K$, respectively. These figures agree with the spectral  type for the primary we derive from an analysis of archival IUE spectra, viz.\, B2--3\,III. A detailed study of V\,393 Sco is in preparation.

\vspace{1cm}

We acknowledge support by projects Fondecyt 1070705, Fondecyt 3085010,  FONDAP 15010003, BASAL PFB-06, MECESUP and  Sociedad Chilena de Astronom\'{\i}a.

\end{document}